\documentclass[12pt]{article}
\usepackage{amsfonts,amsmath,amssymb,authblk,color,enumerate,float,graphicx,multicol,multirow,placeins,pifont,slashed,titlesec,xspace}
\usepackage[colorlinks=true,urlcolor=blue,anchorcolor=blue,citecolor=blue,filecolor=blue,linkcolor=blue,menucolor=blue,pagecolor=blue,linktocpage=true]{hyperref}
\usepackage[compress,numbers]{natbib}
\usepackage[labelfont=bf]{caption}
\usepackage[top=1in,bottom=1in,left=1in,right=1in,footskip=0.5in,headheight=0.5in,footnotesep=0.2in,marginparwidth=0in,marginparsep=0in]{geometry}
\usepackage{tikzfeynman}

\renewcommand{\arraystretch}{1.5}
\long\def\symbolfootnote[#1]#2{\begingroup%
\def\thefootnote{\fnsymbol{footnote}}\footnote[#1]{#2}\endgroup}

\newcommand{\abs}[1] {\big | #1 \big |}
\newcommand{\xmark}  {\ding{55}}%
\newcommand\notype[1]{\unskip}

\begin{document}
\begin{flushright}
\small  CERN-TH-2018-004
\end{flushright}
\vspace{.6cm}

\begin{center}
\pagenumbering{gobble}
{\LARGE \bf Charged Fermions Below 100 GeV}
\bigskip\vspace{1cm}{

\large
Daniel Egana-Ugrinovic$^1$, Matthew Low$^2$, and Joshua T. Ruderman$^{3,4}$ }
\\[7mm]
 {\small
 $^1$\,C. N. Yang Institute for Theoretical Physics, Stony Brook, NY 11794, USA \\
$^2$\,School of Natural Sciences, Institute for Advanced Study,\\ Einstein Drive, Princeton, NJ 08540, USA\\
$^3$\,Center for Cosmology and Particle Physics, Department of Physics, \\ 
New York University, New York, NY 10003, USA\\
$^4$\,Theoretical Physics Department, CERN, Geneva, Switzerland
 }

\end{center}

\vspace{4ex}
\begin{abstract}
How light can a fermion be if it has unit electric charge?  We revisit the lore that LEP robustly excludes charged fermions lighter than about 100 GeV\@.  We review LEP chargino searches, and find them to exclude charged fermions lighter than 90 GeV, assuming a higgsino-like cross section.  However, if the charged fermion couples to a new scalar, destructive interference among production channels can lower the LEP cross section by a factor of 3.  In this case, we find that charged fermions as light as 75 GeV can evade LEP bounds, while remaining consistent with constraints from the LHC\@.  As the LHC collects more data, charged fermions in the $75-100$ GeV mass range serve as a target for future monojet and disappearing track searches.
\end{abstract}

\newpage
\pagenumbering{arabic}
\setcounter{page}{2}
\tableofcontents
\newpage

\section{Introduction}

New fermions with electroweak quantum numbers may be present at the TeV scale.  New electroweak fermions are motivated  by solutions to the naturalness problem of the Higgs, such as the charginos predicted by supersymmetry.  In addition, weak scale dark matter may be associated with new electroweak fermions~\cite{Lee:1977ua,Gunn:1978gr,Ellis:1983ew,Jungman:1995df,Cirelli:2005uq}, and if these fermions couple to the Higgs, they may also be relevant for understanding the electroweak phase transition \cite{Carena:2004ha,Davoudiasl:2012tu,Egana-Ugrinovic:2017jib}. 

The LHC is now sensitive to charginos with masses of hundreds of GeV~\cite{Aaboud:2017nhr,Aaboud:2017leg,ATLAS:2017uun,Sirunyan:2017zss,Sirunyan:2017obz,Sirunyan:2017lae,Sirunyan:2017qaj,CMS:2017fij}.  As more data are collected, heavier states will come into reach.  Significant attention has been devoted towards maximizing the mass reach for new electroweak fermions (see for example Refs.~\cite{Buckley:2009kv,Lisanti:2011cj, Schwaller:2013baa, Low:2014cba, Han:2015lma, Barducci:2015ffa, Arbey:2015hca, Badziak:2015qca, Bramante:2015una, Baer:2016usl, Arina:2016rbb, Mahbubani:2017gjh, Baer:2017gzf, Curtin:2017bxr,Halverson:2014nwa,Fukuda:2017jmk,Falkowski:2013jya,Krall:2017xij}).  However, it is also important to be mindful of any gaps in the exclusion limits at lower masses, to make sure that new physics is not missed during the march to higher masses.

The conventional view is that LEP sets the strongest bounds on electroweak fermions lighter than about 100 GeV, while the LHC probes higher masses.  The clean environment of LEP, plus the robust production mechanism through $s$-channel $\gamma/Z$, implies powerful bounds on fermions lighter than half the maximum center-of-mass energy, $\sqrt s = 209~\mathrm{GeV}$\@.  Photons from initial state radiation (ISR) can be tagged, and therefore detection does not require energetic decay products from the new fermions.
There has developed a sort of {\it folk bound}: it is commonly believed that  LEP robustly excludes fermions with unit charge lighter than about 100 GeV\@.  Indeed, most LHC searches for charginos only display bounds above 100 GeV~\cite{Aaboud:2017nhr,ATLAS:2017uun,Sirunyan:2017zss,Sirunyan:2017obz,Sirunyan:2017lae,Sirunyan:2017qaj,CMS:2017fij,Aad:2011cwa,Aad:2013yna,Aad:2014gfa,ATLAS:2014fka,Aad:2015jqa,Aad:2015eda,Khachatryan:2014qwa,CMS:2015loa,Khachatryan:2015pot,Khachatryan:2016hns},\footnote{A few LHC searches, mainly from $\sqrt s = 7~\mathrm{TeV}$, do show bounds on masses below 100 GeV~\cite{Aaboud:2017leg,Aad:2012jja,Aad:2012pxa,Aad:2012hba,ATLAS:2012jp,Chatrchyan:2012pka}.} and many theory studies of charginos also only consider masses above 100 GeV.   

In this work, we explore fermions with unit electric charge and masses between $m_Z/2$ and $100~\textrm{GeV}$\@. We limit ourselves to study charged fermions in the higgsino gauge representation and call these fermions charginos, both if they are part of the MSSM or if they arise in a more generic, possibly non-supersymmetric, context. We critically examine the robustness of the LEP-II bounds on charginos.  We do not consider the region below $m_Z/2$, where the $Z$ boson can decay to the new fermions, leading to powerful constraints from LEP-I~\cite{Decamp:1991uy,Abreu:1990qu,Acciarri:1995vf,Alexander:1996ne}.

The LEP SUSY working group combined limits from the 4 experiments: ALEPH, DELPHI, L3, and OPAL (``ADLO'' combination)\@.  A combined limit excludes charginos with masses below 103.5~GeV~\cite{lepsusyLarge}, however this bound relies on a restrictive set of assumptions: (1)~wino-like production cross section, (2)~gaugino-unified relation between the bino and wino mass (implying a large splitting between the chargino and lightest neutralino), and (3)~decoupled sneutrinos.   There is also a combined limit applied to the regime of small splitting between the chargino and neutralino~\cite{lepsusy}, excluding charginos lighter than 92.4~GeV (91.9~GeV) for higgsino (wino)-like cross sections, respectively.  In this work, we review the LEP bounds, including newer searches not included in the LEP SUSY working group combination, and we  find that charginos with a higgsino-like cross section are excluded in the range $m_Z/2$ to approximately 90 GeV.  

\begin{figure}[t!]
\begin{center}
\includegraphics[width=10cm]{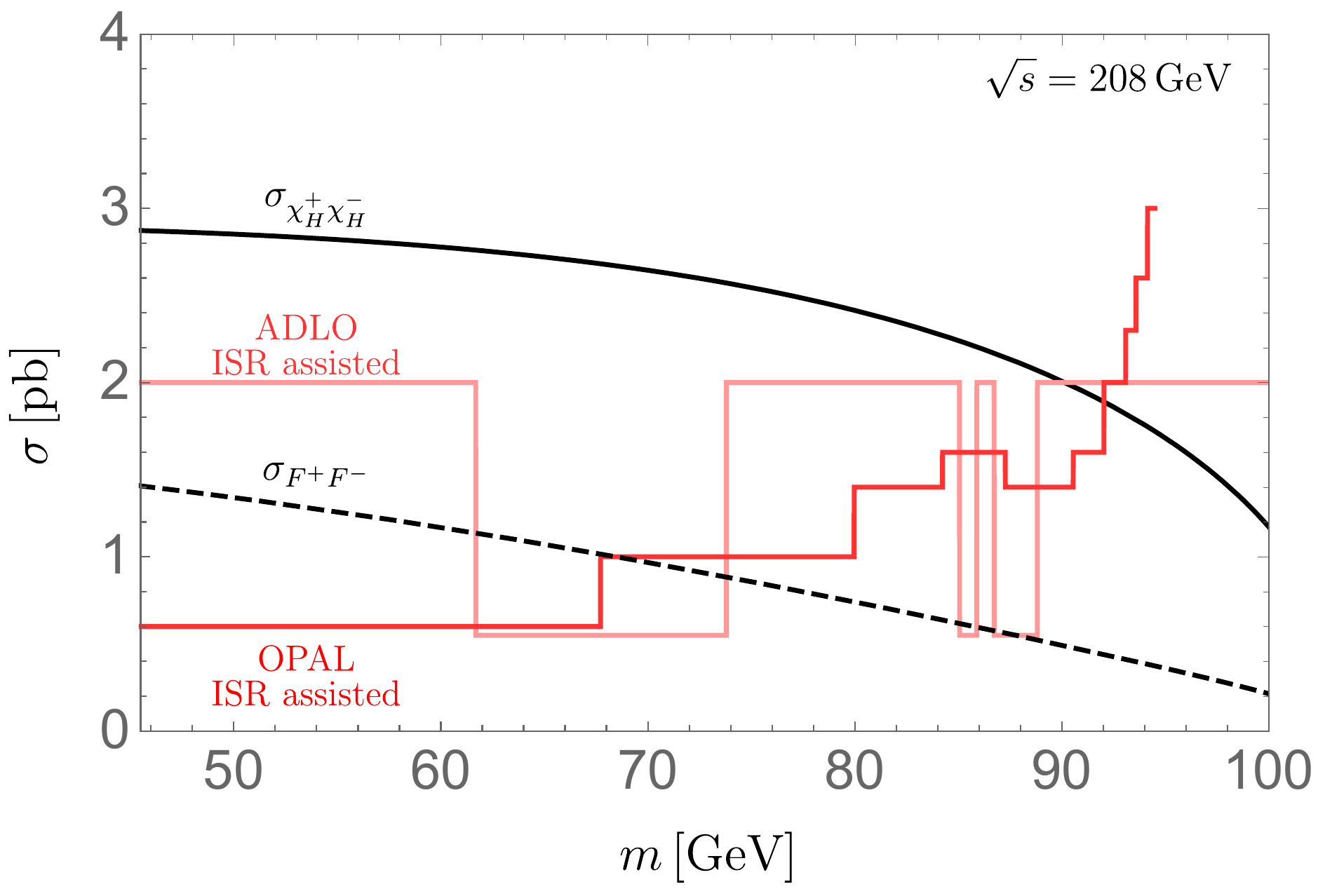}
\caption{LEP higgsino pair production cross section, $\sigma_{\chi_H^+\chi_H^-}$ (black solid), and charged fermion pair production cross section, $\sigma_{F^+F^-}$, allowing for $t$-channel interference (black dashed). In the $t$-channel diagram shown in Fig.~\ref{fig:pairproduction}, the coupling is set to $\kappa=0.5$ and the singlet mass to $m_S = 110~\textrm{GeV}$\@.  The light and dark red lines are the LEP limits on the pair production cross section from~\cite{lepsusy} and~\cite{Abbiendi:2002vz}, respectively. The limits in this figure assume higgsino-like decays, and a charged-neutral mass splitting of $\Delta m = 2.7~\textrm{GeV}$\@.}
\label{fig:higgsinoxsec}
\end{center}
\end{figure}

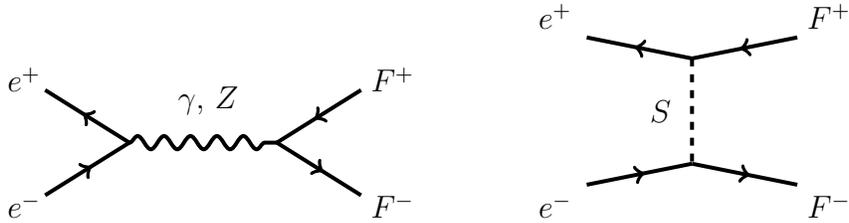
\begin{figure} [t!]
\begin{center}
\begin{tikzpicture}[line width=1.5 pt, scale=1.4]
  \draw[fermionbar]    (-0.8, 0.5)--(0, 0.0);
  \draw[fermion]       (-0.8,-0.5)--(0, 0.0);
  \draw[vector]        (0, 0.0)   --(1.4, 0.0);
  \draw[fermionbar]    ( 1.4, 0.0)--(2.2, 0.5);
  \draw[fermion]       ( 1.4, 0.0)--(2.2,-0.5);
  \node at (0.75, 0.4) {$\gamma$, $Z$};
  \node at (-1.0, 0.6) {$e^+$};
  \node at (-1.0,-0.6) {$e^-$};
  \node at ( 2.5, 0.6) {$F^+$};
  \node at ( 2.5,-0.6) {$F^-$};
\end{tikzpicture}
\quad\quad\quad 
\begin{tikzpicture}[line width=1.5 pt, scale=1.4]
  \draw[fermionbar]    (-1, 1.2)--(0, 1.0);
  \draw[fermionbar]    ( 0, 1.0)--(1, 1.2);
  \draw[scalarnoarrow] (0, 1.0) --(0, 0.0);
  \draw[fermion]       (-1,-0.2)--(0, 0.0);
  \draw[fermion]       ( 0, 0.0)--(1,-0.2);
  \node at (-1.3, 1.4) {$e^+$};
  \node at ( 1.3, 1.4) {$F^+$};
  \node at (-1.3,-0.4) {$e^-$};
  \node at ( 1.3,-0.4) {$F^-$};
  \node at (-0.3, 0.5) {$S$};
\end{tikzpicture}
\caption{Feynman diagrams for $F^+ F^-$ pair production at LEP\@.  The $s$-channel production is fixed by the fermion's quantum numbers, while the $t$-channel production depends on the scalar's mass and the strength of the Yukawa interaction between the charged fermions and the scalar.}
\label{fig:pairproduction}
\end{center}
\end{figure}

In Fig.~\ref{fig:higgsinoxsec}, we compare the cross section of higgsino-like charginos to two bounds from LEP\@.  Note that although we find a bound of about 90 GeV, the higgsino cross section is always within an order one factor of the bound for all masses above $m_Z/2$\@.  If the cross section can be reduced, the bound may be significantly weakened.  We consider a deformation of the minimal model, illustrated in Fig.~\ref{fig:pairproduction}, where the cross section is reduced.  A new scalar has a Yukawa coupling with the charged fermion and the electron, such that $t$-channel exchange of the scalar destructively interferes with the usual $s$-channel diagram.  The dashed curve in Fig.~\ref{fig:higgsinoxsec} shows how the fermion's cross section is reduced, for a particular choice of the scalar mass and the strength of its Yukawa interaction.

As we will describe below, we find that the $t$-channel scalar can reduce the LEP-II limit on charginos from about 90 GeV to about 75 GeV\@.  We find that this gap survives current searches at the LHC\@.  A previous study identified a window of light charginos that decay to leptons through displaced vertices and survive LEP searches~\cite{Batell:2013bka}, however this gap has now been closed by LHC searches for displaced leptons~\cite{Aad:2015rba}.  Previous studies have considered charged scalars, below 100 GeV, that are not excluded by LEP~\cite{Pierce:2007ut,Cao:2017ffm}.

The rest of this paper is organized as follows.  In Sec.~\ref{sec:higgsinos}, we review LEP searches and find that a charged fermion, with a higgsino-like cross section, is bounded to be heavier than about 90 GeV\@.  In Sec.~\ref{sec:model}, we introduce the simplified model that we use to study the effect of $t$-channel interference on the chargino bounds.  In Sec.~\ref{sec:leplimits}, we discuss LEP-II limits in the presence of $t$-channel interference.  Then, in Sec.~\ref{sec:LHC}, we evaluate the limits from LHC searches for monojets, multileptons, and disappearing tracks.  Sec.~\ref{sec:conc} contains our conclusions.  We include an appendix that describes the validations of our simulations for recasting LHC searches.

\FloatBarrier
\section{A Review of LEP Limits on Higgsinos} \label{sec:higgsinos}

We start by providing a brief summary of LEP limits on the pure higgsino model. The pure higgsino model corresponds to an extension of the Standard Model with a vector-like pair of color-neutral, $SU(2)_W$ doublet fermions with hypercharge $Y=\pm 1/2$\@.  We assume that discrete symmetries prevent mixing between the new doublet fermions and Standard Model leptons.  At the renormalizable level, all of the new doublet fermion interactions are set by their quantum numbers.

The spectrum of the higgsino system contains one charged and one neutral Dirac fermion: the chargino, $\chi_H^\pm$, and the neutralino, $\chi_H^0$\@.  At dimension-five, the masses of $\chi_H^\pm$ and $\chi_H^0$ may be split by the Weinberg operator (in the MSSM this mass splitting arises from the mixing among the higgsinos, winos, and bino).  At one-loop, there is an additional irreducible contribution to the mass splitting from infrared effects.  In the range $50~\textrm{GeV} \leq m_{\chi_H^\pm} \leq 100~\textrm{GeV}$, this radiative splitting monotonically increases from $206~\textrm{MeV}$ to $256~\textrm{MeV}$~\cite{Thomas:1998wy}.  In this work, we assume that the neutral fermion is the lightest component of the doublet and is stable. 

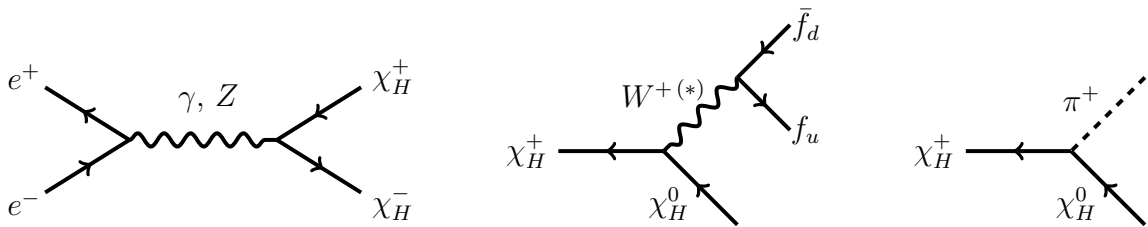
\begin{figure} [h!]
\begin{center}
\begin{tikzpicture}[line width=1.5 pt, scale=1.4]
  \draw[fermionbar]    (-0.8, 0.5)--(0, 0.0);
  \draw[fermion]       (-0.8,-0.5)--(0, 0.0);
  \draw[vector]        (0, 0.0)--(1.4, 0.0);
  \draw[fermionbar]    ( 1.4, 0.0)--(2.2, 0.5);
  \draw[fermion]       ( 1.4, 0.0)--(2.2,-0.5);
  \node at (0.75, 0.4) {$\gamma$, $Z$};
  \node at (-1.0, 0.6) {$e^+$};
  \node at (-1.0,-0.6) {$e^-$};
  \node at ( 2.5, 0.6) {$\chi_H^+$};
  \node at ( 2.5,-0.6) {$\chi_H^-$};
\end{tikzpicture}
\quad\quad 
\begin{tikzpicture}[line width=1.5 pt, scale=1.4]
  \draw[fermion]       (1.0, 0.0)--(0.0, 0.0);
  \draw[fermion]       (1.7, -0.7)--(1.0, 0.0);
  \draw[vector] (1.7, 0.7)--(1.0, 0.0);
     \draw[fermion]       (1.7, 0.7)--(2.2, 0.2);
    \draw[fermion]      (2.2, 1.2)--(1.7, 0.7);
  \node at (-0.3,0.0)  {$\chi_H^+$};
  \node at (1.05,-0.5) {$\chi_H^0$};
  \node at (1,0.55)    {$W^{+\, (*)}$};
    \node at (2.35, 0.2)    {$f_u$};
        \node at (2.35, 1.2)    {$\bar{f_d}$};
\end{tikzpicture}
\quad\quad 
\begin{tikzpicture}[line width=1.5 pt, scale=1.4]
  \draw[fermion]       (1.0, 0.0)--(0.0, 0.0);
  \draw[fermionbar]    (1.0, 0.0)--(1.7, -0.7);
  \draw[scalarnoarrow] (1.7, 0.7)--(1.0, 0.0);
  \node at (1.1,0.5)   {$\pi^+$};
  \node at (1.05,-0.5) {$\chi_H^0$};
  \node at (-0.3,0.0)  {$\chi_H^+$};
\end{tikzpicture}
\caption{Feynman diagrams for charged higgsino pair production at LEP (left) and charged higgsinos decays (center and right).   $f_u,f_d$ stand for Standard Model up or down type quarks or leptons. Both the cross section and branching ratios are fixed by the higgsino's quantum numbers and the chargino and neutralino masses.}
\label{fig:charginodiagrams}
\end{center}
\end{figure}

At LEP, charginos are pair produced via $s$-channel diagrams mediated by gauge bosons and they decay through $W$ bosons into quarks, leptons or pions, as shown in Fig.~\ref{fig:charginodiagrams}. The decays may be two or three-body, depending on the chargino-neutralino mass splitting. Since all interactions are fixed by the higgsino's quantum numbers, the properties of the higgsino system, including the chargino lifetime and branching ratios, are completely determined by the chargino and neutralino masses.

\setlength\tabcolsep{1pt} 
\begin{table} [h!]
\begin{center}\begin{tabular}{|c||c|c|c|c|} \hline
 Search &  ~\textrm{Prompt decays}~ & \textrm{Displaced decays} & $\Delta m \equiv m_{\chi_H^\pm}- m_{\chi_H^0}$ \\ \hline\hline
\begin{tabular}{c}
  \textrm{ADLO conventional} \\
  [-2ex] \cite{lepsusyLarge} (\cite{Jakobs:2001yg})       
\end{tabular}
& 
\begin{tabular}{c}
$\ell,j$, $\slashed{E}_T$
\end{tabular}
&
\xmark
&
$ \geq 3~\textrm{GeV}$ \\  \hline
\begin{tabular}{c}
\textrm{OPAL multilepton}~\cite{Abbiendi:2003ji} 
\end{tabular}
&
\begin{tabular}{c}
  $\ell$, $\slashed{E}_T$
\end{tabular}
 &
 \xmark
 &
 $ \geq 3~\textrm{GeV}$ \\ \hline
\begin{tabular}{c}
  \textrm{ADLO ``low DM''} \\
  [-2ex] \cite{lepsusy} (\cite{Abdallah:2003xe,Heister:2002mn,Acciarri:1999xb,OPALPN464,Chemarin:571780})    
\end{tabular}
& 
\begin{tabular}{c}
$\ell,j$, $\slashed{E}_T$,
\\
[-2ex]
ISR assisted
\end{tabular}
&
\begin{tabular}{c}
kinked tracks,  
\\
[-2ex]
impact parameter 
\end{tabular}
&
\begin{tabular}{c}
$320~\textrm{MeV} \leq \Delta m \leq 10~\textrm{GeV}$ (prompt)
\\
[-2ex]
$m_{\pi^\pm} \leq \Delta m \leq 320~\textrm{MeV} $ (displaced)
\end{tabular}
\\
\hline
\begin{tabular}{c}
  \textrm{OPAL ISR assisted}~\cite{Abbiendi:2002vz}
\end{tabular}
 & 
ISR assisted
&
\xmark
&
$320~\textrm{MeV} \leq \Delta m \leq 5~\textrm{GeV}$ \\ \hline
\begin{tabular}{c}
  OPAL HSCP~\cite{Abbiendi:2003yd} 
\end{tabular}
&
\xmark
&
HSCP
&
$ \leq m_{\pi^\pm}$ \\ \hline
\end{tabular}\end{center}
\caption{Summary of LEP limits used in this work, and the topologies considered within each reference.  Above, $\ell$ stands for leptons $(e,\mu,\tau)$, $j$ for jets, ISR for initial state radiation (of a photon), and HSCP for heavy stable charged particles.  The searches included in the ADLO combination are specified in parentheses.  They include a combination of the analyses using a subset of the full dataset, up to 2001, for conventional searches and using the full dataset, up to 2002, for ``low DM'' compressed searches.  We further break down the ADLO compressed searches into ``prompt'' and ``displaced'' depending on the chargino-neutralino mass splitting, which fixes the chargino lifetime. The ADLO combination also reports limits on HSCP searches for $\Delta m \leq m_{\pi^\pm}$, which we do not use in this work.  Instead, we recast the OPAL HSCP search~\cite{Abbiendi:2003yd}, which provides stronger limits.  None of the OPAL searches in the table are included in the ADLO combination, and they are all performed with the full luminosity.  The last column indicates the charged-neutral higgsino mass splitting covered by each reference.}
\label{tab:LEPsearches}
\end{table}

LEP performed several searches for charged higgsinos heavier than half the $Z$ boson mass. The searches may be divided into different categories depending on the chargino-neutralino mass splitting
\begin{equation}
\Delta m \equiv m_{\chi_H^\pm} - m_{\chi_H^0},
\end{equation}
which controls both the typical momentum of the final state particles and the chargino lifetime~\cite{Abdallah:2003xe}.  The region $\Delta m > 3~\textrm{GeV}$ is covered by conventional searches looking for charginos promptly decaying into leptons and jets. 
For $320~\textrm{MeV} < \Delta m < 3~\textrm{GeV}$, the most effective searches require a photon from ISR as well as other detector activity from the chargino's decay products.  For $m_{\pi^\pm} < \Delta m < 320~\textrm{MeV}$ the chargino lifetime is greater than $\approx 1~\textrm{cm}$, and dedicated searches for disappearing tracks and large impact parameters set the strongest limits. Finally, for mass splittings below the charged pion mass, $\Delta m < m_{\pi^\pm}$, the chargino quickly becomes collider-stable and is probed by heavy stable charged particle (HSCP) searches.  According to~\cite{lepsusy}, the combination of the above searches leads to a lower limit on the chargino mass of $m_{\chi_H^\pm} > 92.4~\textrm{GeV}$\@.

To understand this limit in more detail, we reanalyze a selection of LEP results, which we summarize in Table~\ref{tab:LEPsearches}.  These results include the ADLO results, which are a combination of the ALEPH, DELPHI, L3, and OPAL limits, and additionally, results published afterwards individually by the OPAL collaboration.  

\begin{figure}[h!]
\begin{center}
\includegraphics[width=16cm]{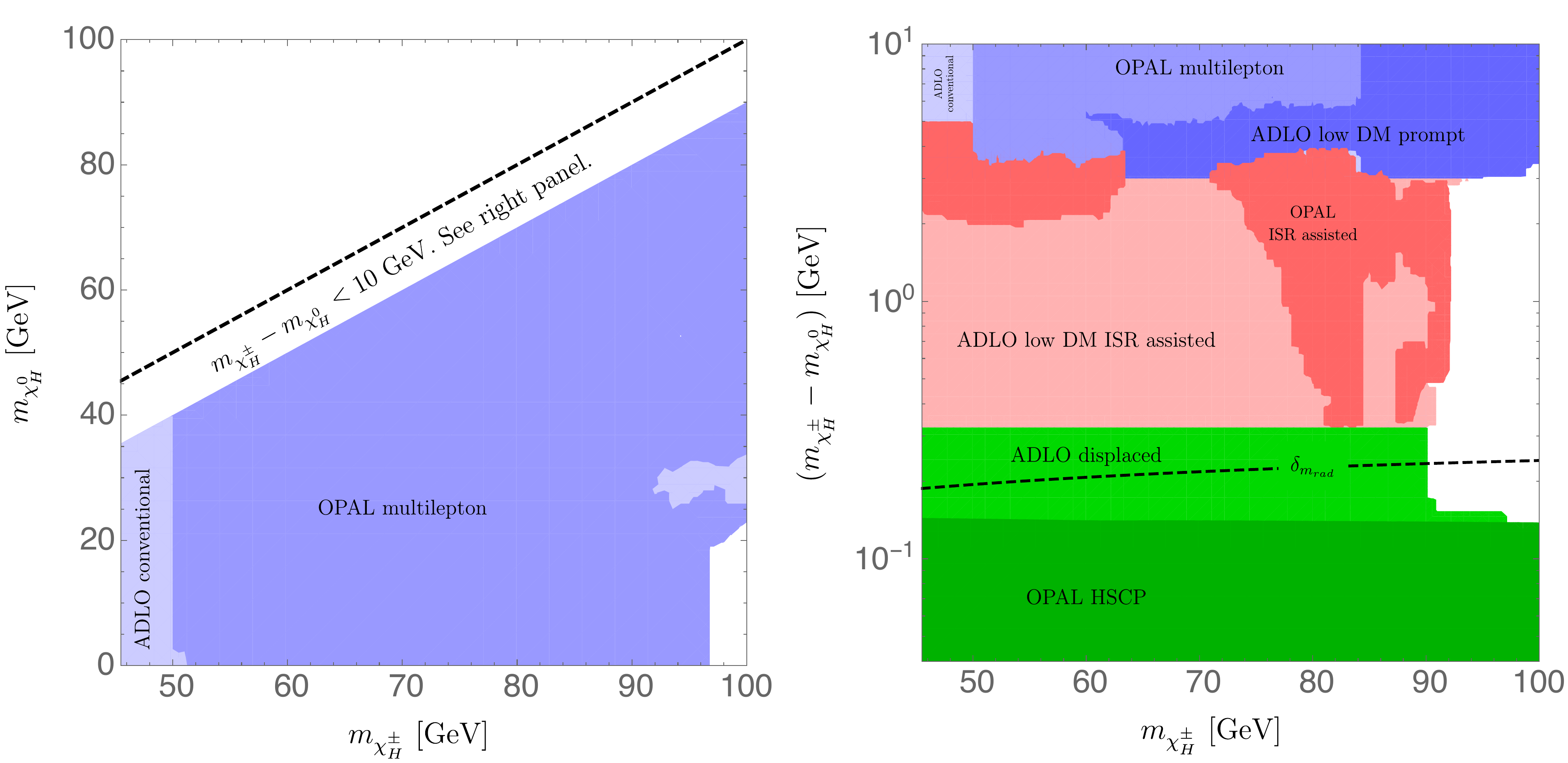}
\caption{Limits on the pure higgsino model, set by the searches in Table~\ref{tab:LEPsearches}, in the uncompressed region (left) and in the compressed region (right).  In the uncompressed region the strongest limits are set by the OPAL multilepton search~\cite{Abbiendi:2003ji} and by ADLO searches for promptly decaying higgsinos~\cite{lepsusyLarge}.   The range $m_{\chi_H^\pm}\leq 50~\textrm{GeV}$ is not covered by the OPAL multilepton search, so we rely exclusively on the ADLO combination.  In the compressed region, the space from $3~\textrm{GeV} \leq \Delta m \leq 10~\textrm{GeV}$ is mainly covered by a combination of the ADLO conventional and ADLO ``low DM'' prompt searches, as well as by the OPAL multilepton search.   The OPAL and ADLO ISR assisted searches~\cite{Abbiendi:2002vz,lepsusy}, set bounds in the regions $320~\textrm{MeV} \leq \Delta m \leq 3~\textrm{GeV}$ and $320~\textrm{MeV} \leq \Delta m \leq 5~\textrm{GeV}$, respectively.  In the region $m_{\pi^\pm} \leq \Delta m \leq 320~\textrm{MeV}$, the chargino decays within the detector, and we rely exclusively on the ADLO combination, which contains dedicated searches for kinked tracks and large impact parameters.  The region $\Delta m \leq  m_{\pi^\pm}$ is covered by the OPAL HSCP search~\cite{Abbiendi:2003yd}, which applies to particles with a decay length $\geq 3~\textrm{m}$\@.  The black dashed line indicates the one-loop radiative chargino-neutralino mass splitting.}
\label{fig:higgsinoplot}
\end{center}
\end{figure}

The results of our analysis are shown in Fig.~\ref{fig:higgsinoplot}.  Most regions of the chargino-neutralino parameter space are excluded by more than one search, so each region is labeled by the search that leads to the strongest limit at that point in parameter space.  We find that charged higgsinos are excluded up to at least $100~\textrm{GeV}$, except in two well-defined regions. In the first region, the mass splitting is large, $\Delta m \gtrsim 60~\textrm{GeV}$\@.  In this case, for $m_{\chi_H^0} \lesssim 25~\textrm{GeV}$, the limit on the chargino mass degrades to $96~\textrm{GeV}$ because the signal kinematics resemble the background from $W$ boson pair production.

The second region with weaker limits occurs when the chargino and neutralino are compressed, but the chargino is not collider-stable, namely $m_{\pi^\pm} \lesssim \Delta m \lesssim 3~\textrm{GeV}$\@.  This region of parameter space is covered by ISR assisted searches and searches for large impact parameters or disappearing tracks.  In this region, the limit on the charged higgsino mass degrades  $\approx 90~\textrm{GeV}$, as discussed in the introduction.  This is the absolute lower limit that we find on the charged higgsino mass, and is approximately consistent with the limit $m_{\chi_H^\pm} > 92.4~\textrm{GeV}$ reported by the ADLO combination~\cite{lepsusy}. 
 
The limits presented in Fig.~\ref{fig:higgsinoplot} rule out most of the parameter space with charginos below $100 \, \textrm{GeV}$, but are specific to the pure higgsino model. In the introduction we pointed out that $\mathcal{O}(1)$ modifications to the pure higgsino production rates may lead to considerably weaker limits.  In the following sections, we investigate quantitatively how the limits on charginos change when the basic assumptions of the pure higgsino model are relaxed.

\section{Simplified Model for Charged Fermions} \label{sec:model}

In this section, we consider a minimal extension of the pure higgsino model to illustrate how simple deviations from this model modify the LEP phenomenology.

We add to the Standard Model a pair of color-neutral, vector-like doublet fermions $F$ and $\bar{F}$ with hypercharges $Y=-1/2$ and $Y=1/2$, respectively, as in the pure higgsino model. The charged and neutral components of the doublets are defined as
\begin{equation}
F =
{\def\arraystretch{1}\tabcolsep=10pt
 \left(\begin{array}{c}
 F^0 \\ F^- 
 \end{array}\right),
}
\quad \quad \quad
\bar{F} 
= 
{\def\arraystretch{1}\tabcolsep=10pt
\left(\begin{array}{c} 
F^+  \\ \bar{F}^{\,0}  
\end{array}\right)
}
.
\end{equation}
We refer to the fermions $F^\pm$ as charginos to indicate that they have unit charge and are part of an $\textrm{SU(2)}_W$ doublet, even though our simplified model is non-supersymmetric. 

Consider now introducing a real scalar singlet $S$, which couples to the doublet $F$ and the electron doublet $L_e$\@.  Up to dimension-five, the Lagrangian contains the operators,
\begin{equation} \label{eq:simplifiedlagrangian1}
- \mathcal{L} \supset
m_F F \bar{F} 
+ \frac{m_S^2}{2}  S^2
+ \frac{1}{\Lambda} (F H) (\bar{F} H^c)
+ \kappa \, L_e \bar{F} S + {\rm h.c.} + V(H,S).
\end{equation}
We require the potential, $V(H,S)$, to be minimized at the origin of the field space of $S$, so that it does not condense.  This model does not violate lepton number, as can be seen by assigning $F$ and $\bar{F}$  electron numbers of  $1$ and $-1$, respectively.  Individual lepton flavor numbers are also preserved, implying that the model is safe from flavor constraints. Additional renormalizable interactions beyond the ones in Eq.~\eqref{eq:simplifiedlagrangian1} are easily forbidden by imposing discrete and continuous global symmetries.  Such symmetries forbid mixing of the new doublets with Standard Model fermions and stabilize the lightest component of the singlet-doublet sector. The only dimension-five term we include is the Weinberg operator, which is responsible for splitting the neutral and charged components of the $SU(2)_W$ doublets at tree-level.\footnote{The Weinberg operator $(F H) (\bar{F} H^c)$ may be obtained from integrating out a heavy complex singlet with electron number coupling to the bilinears $F H$ and $\bar{F} H^c$ at tree-level. Note that in this case the operator $(F\bar{F})(H^\dagger H)$, which may not be forbidden by continuous and discrete symmetries respected by the interactions in Eq.~\eqref{eq:simplifiedlagrangian1}, is not generated.} The couplings $\kappa$ and $\Lambda$ are generically complex, but for simplicity we set the phases to zero and do not study the $CP$-violating phenomenology.  Without loss of generality, we work in the basis where $m_F \geq 0$ and $\kappa \geq 0$\@.

In addition to the scalar singlet with mass $m_S$, the model contains one charged Dirac fermion with mass $m_{F^{\pm}}=m_F$ and one neutral Dirac fermion with mass $m_{F^0}$\@. The mass splitting between the charged and neutral fermions is
\begin{equation} \label{eq:neutralmass}
\Delta m \equiv m_{F^{\pm}}-m_{F^0}
= \frac{v^2}{2 \Lambda} + \delta_{m_{\rm rad}},
\end{equation}
where the Higgs condensate is $v = 246~\textrm{GeV}$ and $\delta_{m_{\rm rad}}$ is positive and accounts for the radiative splitting of the doublet.  We assume that $m_{F^{\pm}}>m_{F^0}$\@.

\begin{figure}[htbp]
\begin{center}
\includegraphics[width=16cm]{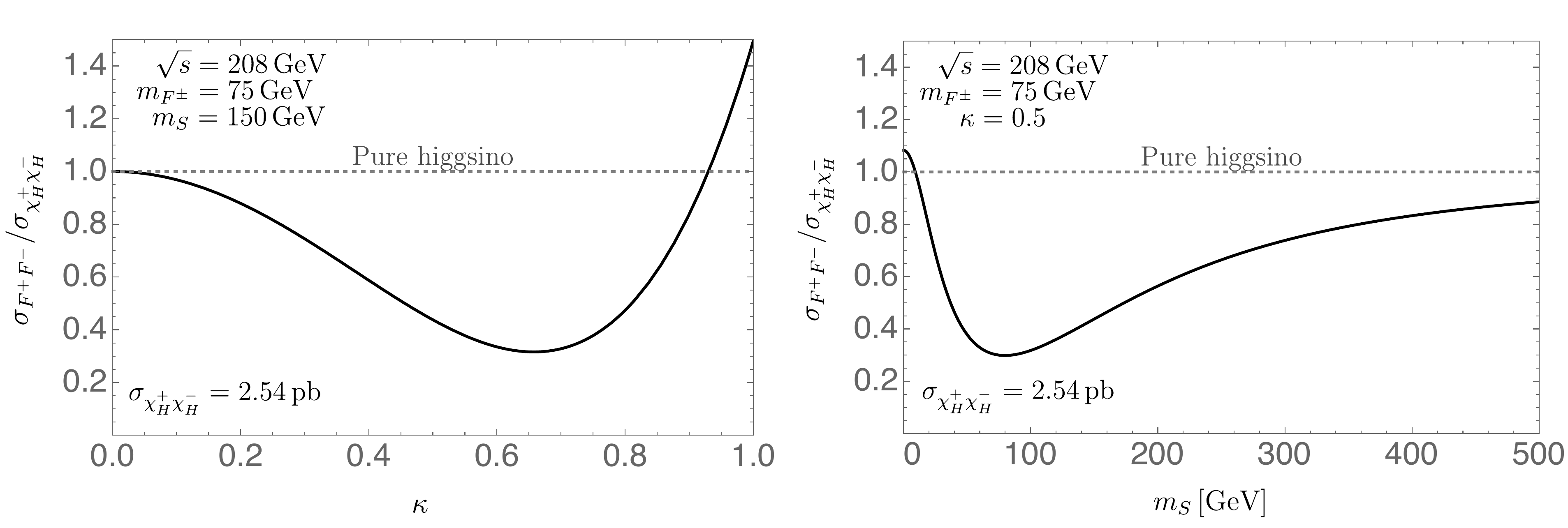}
\caption{Chargino pair production cross section, $\sigma_{F^+F^-}$, at LEP within our simplified model, normalized to the charged higgsino pair production cross section $\sigma_{\chi_H^+\chi_H^-}$.  The cross section is shown as a function of the coupling $\kappa$ for fixed singlet mass $m_S=150~\textrm{GeV}$ (left), and as a function of the singlet mass $m_S$ for fixed coupling $\kappa=0.5$ (right).  In both cases, the charged fermion mass is set to $m_{F^\pm}=75~\textrm{GeV}$\@.  Cross sections are obtained from~\cite{Dreiner:2008tw}.}
\label{fig:fpmxseccombined1}
\end{center}
\end{figure}

\begin{figure}[htbp]
\begin{center}
\includegraphics[width=7.3cm]{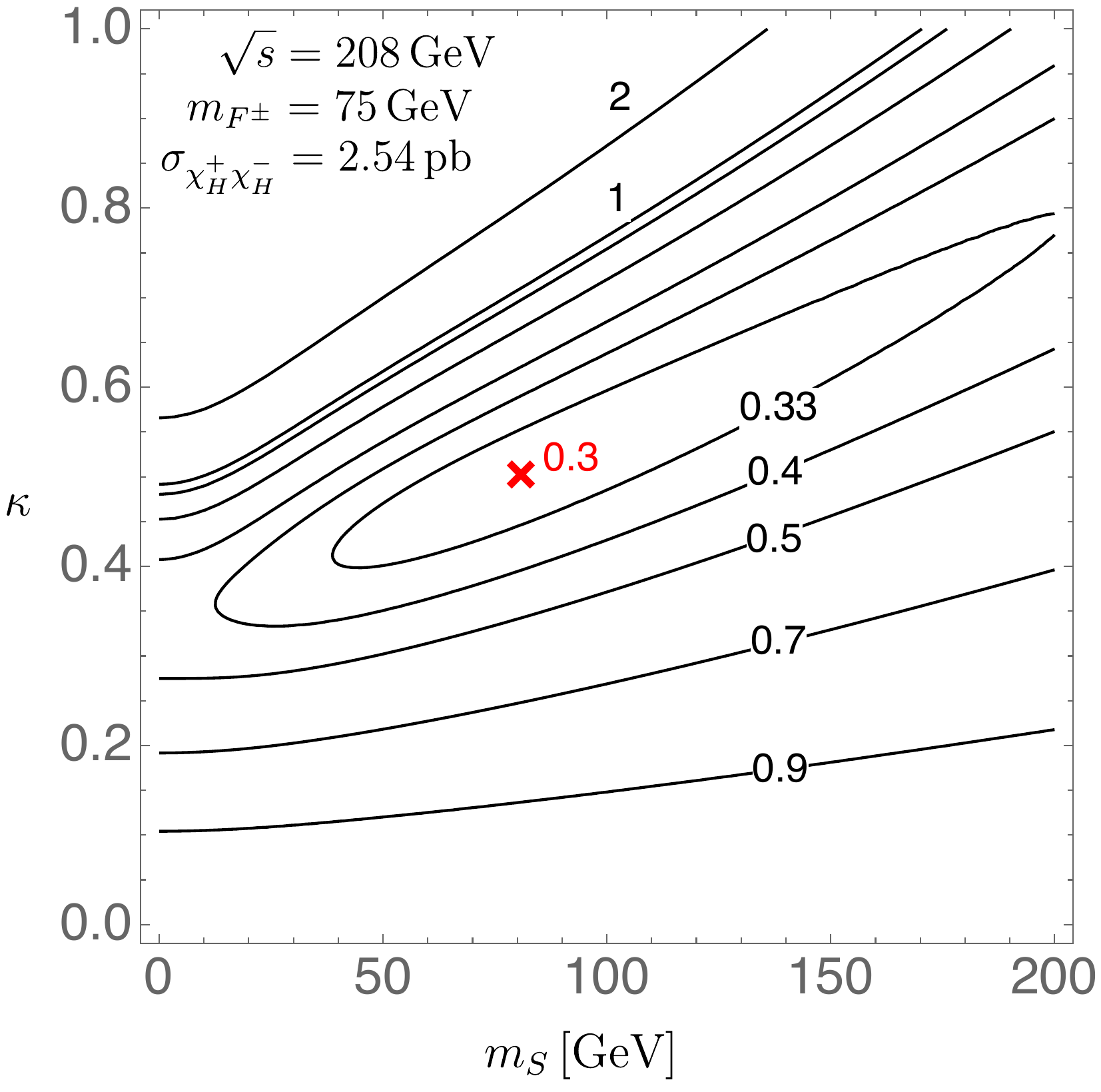}
\caption{Chargino pair production cross section, $\sigma_{F^+F^-}$, at LEP within our simplified model, normalized to the charged higgsino pair production cross section $\sigma_{\chi_H^+\chi_H^-}$.
The cross section is shown as a function of the coupling $\kappa$ and singlet mass $m_S$, for $m_{F^\pm}=75~\textrm{GeV}$\@.  The red cross indicates the point of maximal $s$ and $t$-channel interference, at which the cross section is minimal.  Cross sections are obtained from~\cite{Dreiner:2008tw}.}
\label{fig:fpmxseccombined2}
\end{center}
\end{figure}

The model in Eq.~\eqref{eq:simplifiedlagrangian1} is very similar to the pure higgsino model, but the Yukawa interaction, $\kappa \, L_e \bar{F} S$, leads to two important modifications to LEP phenomenology. First, the pair production rate of $F^\pm$ at LEP is modified with respect to the pure higgsino case, since a new $t$-channel singlet-mediated contribution interferes destructively with the $s$-channel gauge-mediated contribution. The diagrams contributing to the production cross section are shown in Fig.~\ref{fig:pairproduction}.  To show the effect of this interference, in Figs.~\ref{fig:fpmxseccombined1} and~\ref{fig:fpmxseccombined2} we plot the LEP $F^\pm$ pair production cross section normalized to the charged higgsino cross section, as a function of the coupling $\kappa$ and the singlet mass $m_S$ for $m_{F^\pm}=75~\textrm{GeV}$\@.  We see that over a wide range of couplings and masses, the LEP pair production cross section is reduced with respect to the pure higgsino case (which is recovered in the limits $\kappa \rightarrow 0 $ or $m_S \rightarrow \infty$).  For $m_{F^\pm}=75~\textrm{GeV}$, the absolute minimum is obtained for $\kappa=0.5$ and $m_S=81~\textrm{GeV}$, at which point the cross section is reduced to a factor of $0.3$ of the cross section when $\kappa=0$\@.  This minimum is indicated by the red cross in Fig.~\ref{fig:fpmxseccombined2}. 

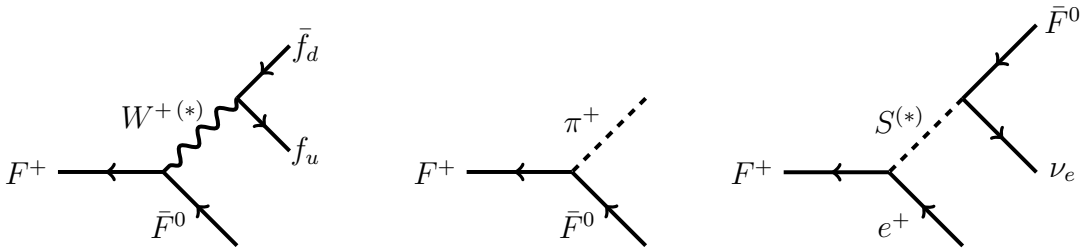
\begin{figure} [H]
\begin{center}
\begin{tikzpicture}[line width=1.5 pt, scale=1.4]
  \draw[fermion]       (1.0, 0.0)--(0.0, 0.0);
  \draw[fermion]       (1.7, -0.7)--(1.0, 0.0);
  \draw[vector]        (1.7, 0.7)--(1.0, 0.0);
       \draw[fermion]       (1.7, 0.7)--(2.2, 0.2);
    \draw[fermion]      (2.2, 1.2)--(1.7, 0.7);
  \node at (-0.3,0.0)  {$F^+$};
  \node at (1.05,-0.5) {$\bar{F}^0$};
  \node at (1,0.55)    {$W^{+\, (*)}$};
      \node at (2.35, 0.2)    {$f_u$};
        \node at (2.35, 1.2)    {$\bar{f_d}$};
\end{tikzpicture}
\quad\quad 
\begin{tikzpicture}[line width=1.5 pt, scale=1.4]
  \draw[fermion]       (1.0, 0.0)--(0.0, 0.0);
  \draw[fermion]       (1.7, -0.7)--(1.0, 0.0);
  \draw[scalarnoarrow] (1.7, 0.7)--(1.0, 0.0);
  \node at (1.1,0.5)   {$\pi^+$};
  \node at (1.05,-0.5) {$\bar{F}^0$};
  \node at (-0.3,0.0)  {$F^+$};
\end{tikzpicture}
\quad\quad 
\begin{tikzpicture}[line width=1.5 pt, scale=1.4]
   \draw[fermion]       (1.0, 0.0)--(0.0, 0.0);
   \draw[fermion]       (1.7, -0.7)--(1.0, 0.0);
   \draw[scalarnoarrow] (1.7, 0.7)--(1.0, 0.0);
   \node at (1.1,0.5)   {$S^{(*)}$};
   \node at (1.05,-0.5) {$e^+$};
   \draw[fermion]       (2.4, 1.4)--(1.7, 0.7);
   \draw[fermion]       (1.7, 0.7)--(2.4, 0.0);
   \node at (-0.3,0.0)  {$F^+$};
   \node at (2.65,1.45) {$\bar{F}^0$};
   \node at (2.65,0)    {$\nu_e$};
\end{tikzpicture}
\caption{Feynman diagrams for charged fermion $F^\pm$ decays. $f_u,f_d$ stand for Standard Model up or down type quarks or leptons. The decays through a $W^*$ (left) or the two body decays into $\pi^\pm$ (center) are set by the fermion's quantum numbers, while the singlet-mediated decay width (right) is controlled by the coupling $\kappa$ and the scalar singlet mass $m_S$\@.  In the singlet-mediated diagram, both $\bar{F}^0 \nu_e$ and $F^0 \bar{\nu}_e$ final states are possible.}
\label{fig:decaymodes}
\end{center}
\end{figure}

\begin{figure}[htbp]
\begin{center}
\includegraphics[width=16cm]{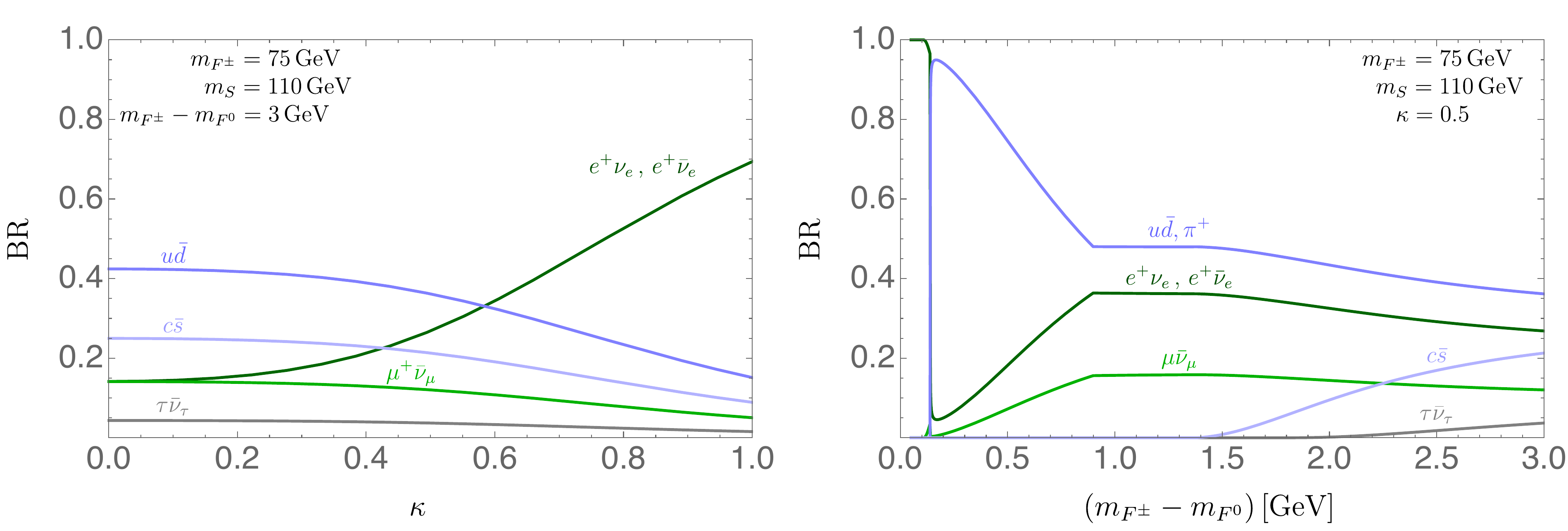}
\caption{Charged fermion $F^\pm$ decay branching fractions as a function of the coupling $\kappa$ for a mass splitting of $m_{F^\pm}-m_{F^0}=3~\textrm{GeV}$ (left), and as a function of the mass splitting for coupling $\kappa=0.5$ (right).  In both cases, the charged fermion mass is $m_{F^\pm}=75~\textrm{GeV}$ and the scalar singlet mass is $m_S=110~\textrm{GeV}$\@.  Decays into the charged pion are matched to decays into up and down quarks at $\Delta m = 0.9~\textrm{GeV}$\@.  Decay widths are obtained from~\cite{Thomas:1998wy,Djouadi:2001fa}.}
\label{fig:BRplots}
\end{center}
\end{figure}

The second effect of the singlet-doublet Yukawa interaction is to alter the decay branching fractions and lifetime of the charged fermion. When $m_S>m_{F^\pm}$, as we assume for the rest of this work, the scalar singlet mediates a new three-body decay mode, shown in Fig.~\ref{fig:decaymodes} (right panel). In Fig.~\ref{fig:BRplots} (left panel), we show the branching fractions into quarks, leptons, and pions as a function of the coupling $\kappa$, and in Fig.~\ref{fig:BRplots} (right panel) as a function of the charged-neutral fermion mass splitting $\Delta m$\@. Since the new Yukawa interaction couples the fermion doublets to electrons, larger values of this interaction increase the branching fraction to electrons.  This modification to the branching ratios results in more electron-rich decays, which alter the LEP search efficiencies relative to the pure higgsino case.  Moreover, the singlet-mediated decay mode increases the charged fermion width (for fixed masses).  This effect is particularly strong for mass splittings below the pion threshold.  For example, for $\Delta m = 100~\textrm{MeV}$ and $m_{F^\pm}=80~\textrm{GeV}$, increasing $\kappa$ from 0 to 1 lowers the decay length from 57 to $3~\textrm{m}$\@.

\subsection*{UV Completions}

We conclude this section by briefly commenting on some possible UV completions of our simplified singlet-doublet model.  One motivation for this is the fact that the electroweak-scale mass of the scalar singlet within our simplified model is not technically natural.  This is easily remedied by, for instance, promoting the scalar to be part of a chiral superfield in a supersymmetric setup.

The singlet-doublet model is a simplified version of the wino-bino-sneutrino system.  In the MSSM with decoupled higgsinos, the lightest chargino is wino-like.  At LEP, $s$-channel production proceeds via gauge-mediated diagrams, while the interfering $t$-channel diagram is mediated by the electron sneutrino.  In this case, in addition to direct limits on charginos, one must also consider the limits on the direct production of the left-handed selectron.  We leave a detailed study of the wino-sneutrino system for future work.

An alternative UV realization of our simplified model is to consider the MSSM extended by a complex scalar singlet charged under electron number. The scalar singlet can then couple to the higgsinos through a superpotential interaction, $\kappa L_e H_u S$, where $H_u$ is the up-type Higgs superfield and $S$ has been promoted to a superfield. In this theory, the charged higgsinos would play the role of the fermions $F^\pm$ in our simplified model.  If the fermion partners of the singlet field are heavy, the discussion would be similar to the one in this work, but with a complex scalar in the effective theory providing the $t$-channel interference in Fig.~\ref{fig:pairproduction} instead of a real scalar.

\section{LEP Limits on the Simplified Model} \label{sec:leplimits}

In the previous section, we found that a simple modification to the pure higgsino benchmark scenario, namely the addition of a singlet scalar, can lead to significant differences in the production rates and decay branching fractions relevant for LEP searches for charginos.  In this section, we reanalyze LEP limits in the context of our simplified model.  Due to the modified branching ratios, the relative composition of final states is different in the simplified model compared to the pure higgsino model.  As a result, the overall search efficiency is different in the simplified model than in the benchmark models considered in the experimental searches.  We take a conservative approach to setting limits meaning that we only set limits from experimental searches that can be reliably recast.  When insufficient information about a search is available we do not set limits using that search, however, we do show the would-be limits under specified assumptions.  We make use of the different searches as follows.

The OPAL multilepton search~\cite{Abbiendi:2003ji} sets bounds on the chargino pair production cross section times branching fraction squared into electrons, muons, and hadronically-decaying taus. The search assumes lepton flavor universality, which is violated in our simplified model since the singlet $S$ mediates three-body decays into electrons only (see Fig.~\ref{fig:decaymodes}).  The efficiencies of electrons and muons are similar, and higher than that of hadronically-decaying taus~\cite{Abbiendi:2003ji}.  Consequently, the search efficiency in our simplified model should be larger than in the flavor universal scenario, since decays to electrons are enhanced.  There is not enough information presented by OPAL to determine the efficiencies for separate final states, so we conservatively apply the OPAL limit by assuming the same overall efficiency for leptonic final states, despite the higher efficiency expected in our simplified model.

The OPAL ISR assisted search~\cite{Abbiendi:2002vz} and ADLO combination with prompt decays~\cite{lepsusyLarge,lepsusy} set limits on the charged fermion pair production cross section assuming three body decays through a $W^{(*)}$ or two body decays into a charged pion, with rates fixed by the higgsino quantum numbers.  Due to $S$-mediated decays into electrons, in our simplified model the $W^{(*)}$ and $\pi$ decay modes are diluted with respect to the pure higgsino case by a common factor, so we simply dilute the reported limits on the cross section by this common factor squared. This choice is again conservative, since it does not take into account the gain in efficiency due to the additional electrons in the final state. 

In the case of the ADLO combination with searches for kinked tracks or for large impact parameters ~\cite{lepsusy}, estimating the efficiencies is more challenging, since they depend on both the decay branching fractions and the chargino lifetime. For these searches, we only present for reference the limits that one would obtain by (crudely) assuming the same efficiencies as in the pure higgsino case.    In Sec.~\ref{sec:LHC}, we recast LHC searches for disappearing tracks to provide a more reliable bound in the case of displaced decays.

Finally, the OPAL HSCP search~\cite{Abbiendi:2003yd} covers the very small mass splitting region with collider-stable charginos.  In this case, we simply use the reported pure higgsino cross section limits by rescaling by the fraction of events where both charged fermions have a flight distance longer than $3~\textrm{m}$, as required by the search. 

\begin{figure}[htbp]
\begin{center}
\includegraphics[width=16cm]{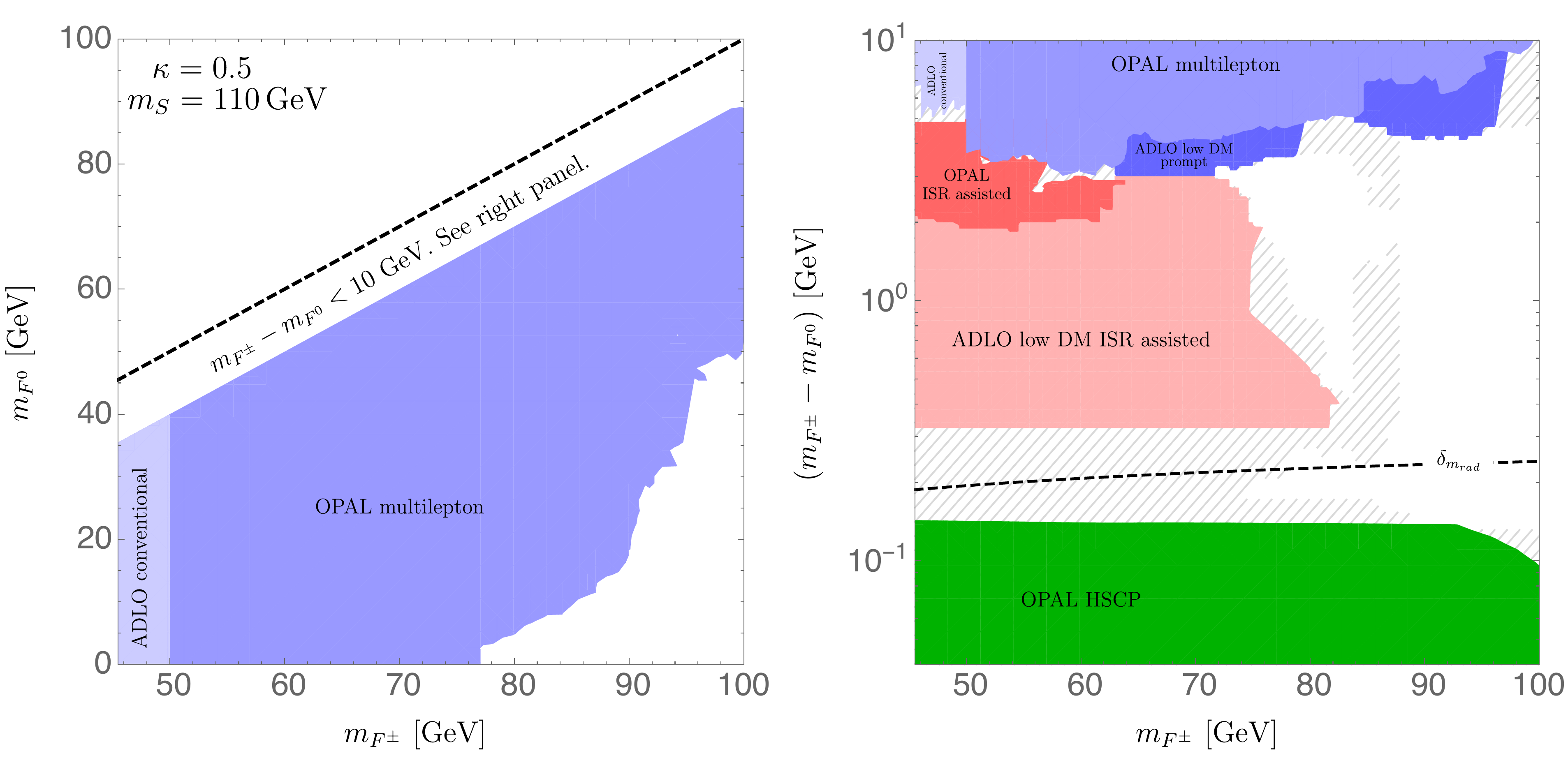}
\caption{Limits on charginos $F^\pm$ (from the simplified model of Sec.~\ref{sec:model}), set by the searches in Table~\ref{tab:LEPsearches}, in the uncompressed region (left) and in the compressed region (right).  The coupling is set to $\kappa=0.5$ and the scalar singlet mass to $m_S=110~\textrm{GeV}$\@.  
All the searches included present limits for the pure higgsino, so we present limits for our simplified model by conservatively estimating the efficiencies, as explained in Sec.~\ref{sec:leplimits}.  For comparison, we also show in hatched regions the excluded space that one would obtain assuming the same efficiencies as in the pure higgsino scenario. 
In the uncompressed region, limits are obtained from the OPAL multilepton search~\cite{Abbiendi:2003ji} and the ADLO combination with promptly-decaying higgsinos~\cite{lepsusyLarge}.  In the compressed region, limits are obtained from the ADLO conventional combination~\cite{lepsusyLarge},  ADLO ``low DM'' combination~\cite{lepsusy}, OPAL multilepton, OPAL ISR assisted, and OPAL HSCP searches~\cite{Abbiendi:2003ji,Abbiendi:2002vz,Abbiendi:2003yd}. The black dashed line indicates the one-loop radiative charged-neutral mass splitting.}
\label{fig:kappa05}
\end{center}
\end{figure}

The resulting limits, for coupling $\kappa=0.5$ and singlet mass $m_S=110~\textrm{GeV}$, are presented in Fig.~\ref{fig:kappa05}, where different colors represent the searches leading to the strongest limit on the pair production cross section at each point.  The hatched regions show the would-be limits by assuming that the efficiencies are unchanged between the pure higgsino model and the simplified model.  

From Fig.~\ref{fig:kappa05} (left panel), we see that in the uncompressed region, $\Delta m \geq 10~\textrm{GeV}$, the absolute LEP limit on the charged fermion mass is $m_{F^\pm} \geq 77~\textrm{GeV}$\@. The weakest end of the limit is achieved for $m_{F^0} \lesssim 5~\textrm{GeV}$\@. In this region, multilepton searches lose sensitivity due similarity between the kinematics of the signal and the $W^+ W^-$ background. 

In Fig.~\ref{fig:kappa05} (right panel), we show the limits in the compressed region, $\Delta m \leq 10~\textrm{GeV}$\@.  First, we note that we cannot set reliable bounds in the region $m_{\pi^\pm} \leq  \Delta m \leq 320~\textrm{MeV}$, due to our inability to reliably recast kinked track and large impact parameter searches at LEP\@.  The hatched region indicates the would-be limit (crudely assuming higgsino-like efficiencies) and rules out charginos up to $\approx 75~\textrm{GeV}$\@.  In the highly compressed region, which is covered by HSCP searches, $\Delta m \leq m_{\pi^\pm}$, the limit on charginos is $m_{F^\pm} \gtrsim 92~\textrm{GeV}$, which is weaker than in the pure higgsino case due to the smaller charged fermion lifetime, as discussed in the previous section.

For promptly decaying fermions, $\Delta m \gtrsim 300~\textrm{MeV}$, we see from Fig.~\ref{fig:kappa05} (right panel) that there are a couple of small gaps in coverage in the range $m_Z/2 \leq m_{F^\pm} \leq 63~\textrm{GeV}$\@.  These gaps are at the interface between the region of validity of different searches. They occur due to unphysical discrete jumps due coarse binning in the excluded cross section reported by the corresponding LEP references, and we expect them to be excluded if more fine-grained limits were provided. In addition, these gaps will be covered by LHC searches (see Sec.~\ref{sec:LHC}).  Disregarding these small gaps, we find that the absolute limit on charginos within our simplified model is $m_{F^\pm} \gtrsim 73~\textrm{GeV}$\@. The weakest end of the limit is achieved for $ \Delta m \approx 2-3~\textrm{GeV}$, a mass splitting region which is covered by ISR assisted searches.  Note that including the hatched regions does not change our conclusions.

Finally, we briefly comment on electroweak precision tests. In the renormalizable theory, since the fermion doublets do not couple to the Higgs, there is no one-loop contribution to the $S$, $T$, and $U$ parameters~\cite{Thomas:1998wy}. On the other hand, at the renormalizable level the extended oblique parameters, $V$, $W$, and $X$, are finite at one loop \cite{Maksymyk:1993zm}. To set limits, we obtain the one-loop $V$, $W$, and $X$ parameters using \textsc{Package-X}~\cite{Patel:2015tea,Patel:2016fam} and perform an electroweak fit as in~\cite{Batell:2013bka}.  We find a $95\%$ CL limit on the charged fermion mass $m_{F^\pm} \geq 54~\textrm{GeV}$, which is independent of coupling $\kappa$ and singlet mass $m_S$ at one-loop. 

In addition to the oblique analysis, we also check the impact on four-fermion operators via the Bhabha scattering, $e^+ e^- \rightarrow e^+ e^-$, cross section at LEP\@.  In our model, the Bhabha scattering cross section is modified at one-loop by box diagrams with $\bar{F}$ and $S$ in the loop.  To set limits, we calculate the amplitudes with \textsc{FeynCalc}~\cite{Mertig:1990an,Shtabovenko:2016sxi} and the one-loop integrals with \textsc{Package-X}\@.  We then perform a full fit to the measured Bhabha scattering cross section for $7$ LEP center of mass energies and $15$ scattering angle bins including full correlations reported in~\cite{Schael:2013ita}.  The limit depends strongly on the singlet-doublet Yukawa coupling and the doublet fermion and scalar singlet masses.  We find that all of the parameter space presented in Fig.~\ref{fig:kappa05} is allowed.  For reference, fixing the fermion doublet masses at $m_{F^\pm}=75~\textrm{GeV}$ and scalar singlet mass $m_S=110~\textrm{GeV}$, we find that couplings of $\kappa \geq 1.5$ are excluded by the precision Bhabha scattering analysis at $95\%$ CL\@.

\section{LHC Limits on the Simplified Model} \label{sec:LHC}

In the previous section, we concluded that LEP rules out charginos within our simplified model with mass $m_{F^\pm}~\leq~77~\textrm{GeV}$ in the uncompressed region, $\Delta m \geq 10~\textrm{GeV}$, and with mass $m_{F^\pm}~\leq~73~\textrm{GeV}$ in the compressed region, $\Delta m \leq 10~\textrm{GeV}$\@.  In this section, we discuss the impact of searches from the LHC\@.

There are a number of searches that can be used to probe charginos at the LHC\@.  Since the fermion doublets $F,\bar{F}$ couple to the Higgs via the Weinberg operator, invisible Higgs decays set constraints which we discuss in Sec.~\ref{sec:higgs}.  In the compressed region, the charged fermions may decay leaving little-to-no activity in the detector, and can be probed by monojet searches, presented in Sec.~\ref{sec:monojets}.  For even smaller mass splittings, the charged fermions may lead to kinked or disappearing tracks, as discussed in Sec.~\ref{sec:dtracks}. Other LHC searches leading to weaker limits are mentioned in Sec.~\ref{sec:otherLHC}.  Finally, the combination of LEP and LHC searches is shown in Sec.~\ref{sec:LEPLHC}.

\subsection{Invisible Decays of the Higgs} \label{sec:higgs}

The Weinberg operator in Eq.~\eqref{eq:simplifiedlagrangian1} leads to an effective dimension-five coupling between the Higgs and the neutral fermions $F^0, \bar{F}^0$
\begin{equation} \label{eq:ghiggs}
g_{hF^0\bar{F}^0} = -\frac{ v}{\Lambda}  = -\frac{2 \Delta m}{v},
\end{equation}
where the second equality uses Eq.~\eqref{eq:neutralmass} at tree-level.  The charged fermion does not couple to the Higgs. The invisible decay width of the Higgs to the neutral fermion is 
\begin{equation}
\Gamma= \frac{m_h}{8\pi} 
g_{hF^0\bar{F}^0}^2 
\left(1-
\frac{4m_{F^0}^2}{m_h^2}
\right)^{3/2}.
\end{equation}
The current limit on the Higgs invisible width is $\textrm{BR}_{h\rightarrow \textrm{inv}} \leq 0.24$~\cite{Aad:2015pla,Khachatryan:2016whc}, and rules out most of the parameter space with $m_{F^0}\leq m_h/2$, except when $m_{F^\pm}$ and $m_{F^0}$ are very compressed (since the compression also suppresses the couplings to the Higgs).  The corresponding limits are shown in Fig.~\ref{fig:LEPLHC}.

\subsection{Monojet Searches} \label{sec:monojets}

In the compressed region of $300~\textrm{MeV} \leq \Delta m \leq 10~\textrm{GeV}$, the decay products of the charged fermion are soft and therefore challenging to detect at the LHC\@.  In this region, fermion pair production ($F^+ F^-$, $F^+ F^0$, $F^- \bar{F}^0$, and $F^0 \bar{F}^0$) results in a signal with missing energy and little-to-no hadronic or leptonic activity in the detector.  This topology is constrained by LHC searches that look for large missing energy along with an ISR jet, namely monojet searches. For splittings below $ \sim 300~\textrm{MeV}$ the decays are no longer prompt, and dedicated searches for displaced objects  become effective.

There are 13 TeV monojet searches from both ATLAS~\cite{Aaboud:2017phn} and CMS~\cite{CMS:2017tbk} with about $36~\textrm{fb}^{-1}$ of data.  Here, for simplicity, we just recast the CMS search, which is representative of both (but sets stronger limits due to an apparent downward fluctuation). This search selects events by defining 22 exclusive $\slashed{E}_T$ regions, from $\slashed{E}_T = 250$ to  $1400~\textrm{GeV}$\@.  The leading jet is required to have a transverse momentum of $p_T\geq 100~\textrm{GeV}$ and a pseudorapidity of $\abs{\eta}\leq 2.5$\@. A $p_T$ cut on leptons, taus, photons, and $b$-jets is imposed, and minimum angles are required between the four leading jets.

In order to recast the CMS monojet search, we perform a Monte Carlo Simulation.  We implement our simplified model with \textsc{FeynRules} \cite{Alloul:2013bka} and simulate events at leading order with \textsc{Madgraph5\_aMC@NLO} \cite{Alwall:2014hca}, using the \textsc{nn23lo1} PDF dataset \cite{Ball:2013hta}. We use \textsc{Pythia8}~\cite{Sjostrand:2007gs} to simulate the parton shower, and  \textsc{Delphes 3}  to perform the detector simulation using the CMS detector card \cite{deFavereau:2013fsa}. Jets are clustered using the anti-$k_T$ algorithm~\cite{Cacciari:2008gp} with jet radius $R=0.4$\@.  We match up to three jets using the MLM matching scheme \cite{Alwall:2007fs} with a matching scale of $50~\textrm{GeV}$. In Appendix~\ref{app:validation}, we describe the validation of our simulaton.

For the signal, we generate a sample of fermion pair production events.  We set limits using the ${\rm CL}_s$ method~\cite{Read:2002hq}, and combine the limits from the different missing energy bins by making use of the bin with the best expected limit at each point in the model parameter space.  The resulting limits for coupling $\kappa=0.5$ and singlet mass $m_S=110~\textrm{GeV}$ are presented in Fig.~\ref{fig:LEPLHC}.  The limits are mostly driven by the low missing transverse energy bins, $\slashed{E}_T \leq 590~\textrm{GeV}$, for which the errors are already close to being dominated by systematics.

We note that the limits are roughly independent of $\kappa$\@ in the compressed region. This is because monojets searches are sensitive to prompt decays, and therefore the only effect of the Yukawa coupling is to change fermion branching ratios, to which the monojet searches are not sensitive.

\subsection{Disappearing Track Searches} \label{sec:dtracks}

As the charged-neutral splitting goes below $\sim 300~\textrm{MeV}$, the decay length of the charged fermion becomes macroscopic.  At the LHC, there are a number of searches that target various decays lengths.  Roughly speaking, decay lengths of $\mathcal{O}(\textrm{mm})$ are probed by searches for displaced vertices, $\mathcal{O}(\textrm{cm})$ are probed by searches for kinked or disappearing tracks, and $\mathcal{O}(\textrm{m})$ are probed by searches for heavy stable particles.  
Displaced vertex searches do not set the dominant limit anywhere in our parameter space, because they tend to require energetic particles originating from the displaced vertex~\cite{Aaboud:2017iio}, and are therefore not relevant when the charged and neutral fermions are compressed. Heavy stable particle searches will be discussed in Sec.~\ref{sec:otherLHC}.

The most recent search for disappearing tracks was performed by ATLAS at 13 TeV with $36.1~\textrm{fb}^{-1}$ of data~\cite{Aaboud:2017mpt}.  (The most recent disappearing track search from CMS was at 8 TeV~\cite{CMS:2014gxa}.)  The search looks for the partial track of a chargino, which decays mid-flight to $\chi^0 \pi^\pm$ or $\chi^0 \ell^\pm \nu$\@.  The outgoing pion or lepton is very soft, since its momentum is set by $\Delta m$, and is typically not seen, which means the chargino track appears to end abruptly.  In addition to the disappearing track, this search requires an ISR jet for triggering.

We recast the disappearing tracks search using the following procedure.  Using \textsc{Madgraph5\_aMC@NLO}, we simulate pair production events at leading order and compute the efficiency to select an event with a disappearing track as a function of the lifetime of the charged fermion and its mass (see Appendix~\ref{app:validation} for a more detailed description).  We then compute the expected number of events as a function of lifetime and mass and compare to the 95\% CL excluded number from ATLAS\@.  The results are shown in Fig.~\ref{fig:LEPLHC}.

The disappearing tracks search excludes chargino masses below 100 GeV for mass splittings between 100 and $300~\textrm{MeV}$, for our benchmark point of $m_S=110~\textrm{GeV}$ and $\kappa=0.5$\@.\footnote{The limits do depend on $\kappa$, but only very weakly.  Increasing $\kappa$ leads to a shorter decay length, leading to an overall shift of the limits towards lower mass splittings.  From $\kappa=0$ to 1, the limits on $\Delta m$ only change by $\lesssim 40~\textrm{MeV}$\@.}  For splittings smaller than $100~\textrm{MeV}$, the charged fermion decay length becomes long enough that the majority of charged fermions do not decay within the tracker.  For splittings larger than $300~\textrm{MeV}$, most of the charged fermions decay before they reach the tracker.  Note that $\Delta m = 300~\textrm{MeV}$ corresponds to $c\tau \approx 1~\textrm{cm}$ for our benchmark point, which is naively too short to leave a particle track.  However, the large production cross section and sizable relevant boost factor,  $\beta\gamma \sim 2-5$, imply that limits can be set using the exponential tail of the decay distribution.

\subsection{Other LHC Searches} \label{sec:otherLHC}

As can be seen in Fig.~\ref{fig:LEPLHC}, in the uncompressed region, the invisible Higgs limits close the parameter space left open by LEP while in the compressed region monojet searches, and disappearing track searches work together to constrain some of the parameter space.  There are a number of other searches at the LHC that can be used to constrain the simplified model for chargino masses below $100 \, \textrm{GeV}$\@.  We mention them briefly in this section.

Multilepton searches look for one or more charged leptons. Dedicated searches in the compressed region using an ISR jet have been performed by both ATLAS and CMS~\cite{Aaboud:2017leg,Aad:2015eda,CMS:2017fij}, but still the minimal lepton $p_T$ required in these searches is at least $3.5~\textrm{GeV}$ (for muons at CMS) and more typically $\approx 5 - 10~\textrm{GeV}$\@. For this reason, these searches do not outperform LEP searches for chargino masses below $100~\textrm{GeV}$ and mass splittings below $\approx 3~\textrm{GeV}$\@. This expectation is confirmed by the latest ATLAS results~\cite{Aaboud:2017leg}, which do not improve on the $\approx 90~\textrm{GeV}$ LEP charged higgsino bound at small mass splittings. Multilepton searches at LHC are most effective when the leptons are hard and they set limits in the uncompressed region $\Delta m \geq 10~\textrm{GeV}$,  but as discussed above, the only space uncovered by LEP searches in the uncompressed region is already excluded by invisible Higgs searches.

For $\mathcal{O}(\textrm{cm})$ decay lengths, the disappearing track searches are the most sensitive, while for longer decay lengths, HSCP searches become the most sensitive.  The HSCP searches performed by LEP constrain cross sections at the $\sim 0.01~\textrm{pb}^{-1}$ level, which is far below the cross section in our simplified model within the range of masses that we consider.  Since HSCP searches at the LHC cover approximately the same range of $\Delta m$ as the LEP searches, we do not recast HSCP searches from the LHC\@.

\subsection{Combined LEP and LHC Limits} \label{sec:LEPLHC}

\begin{figure}[htbp]
\begin{center}
\includegraphics[width=16cm]{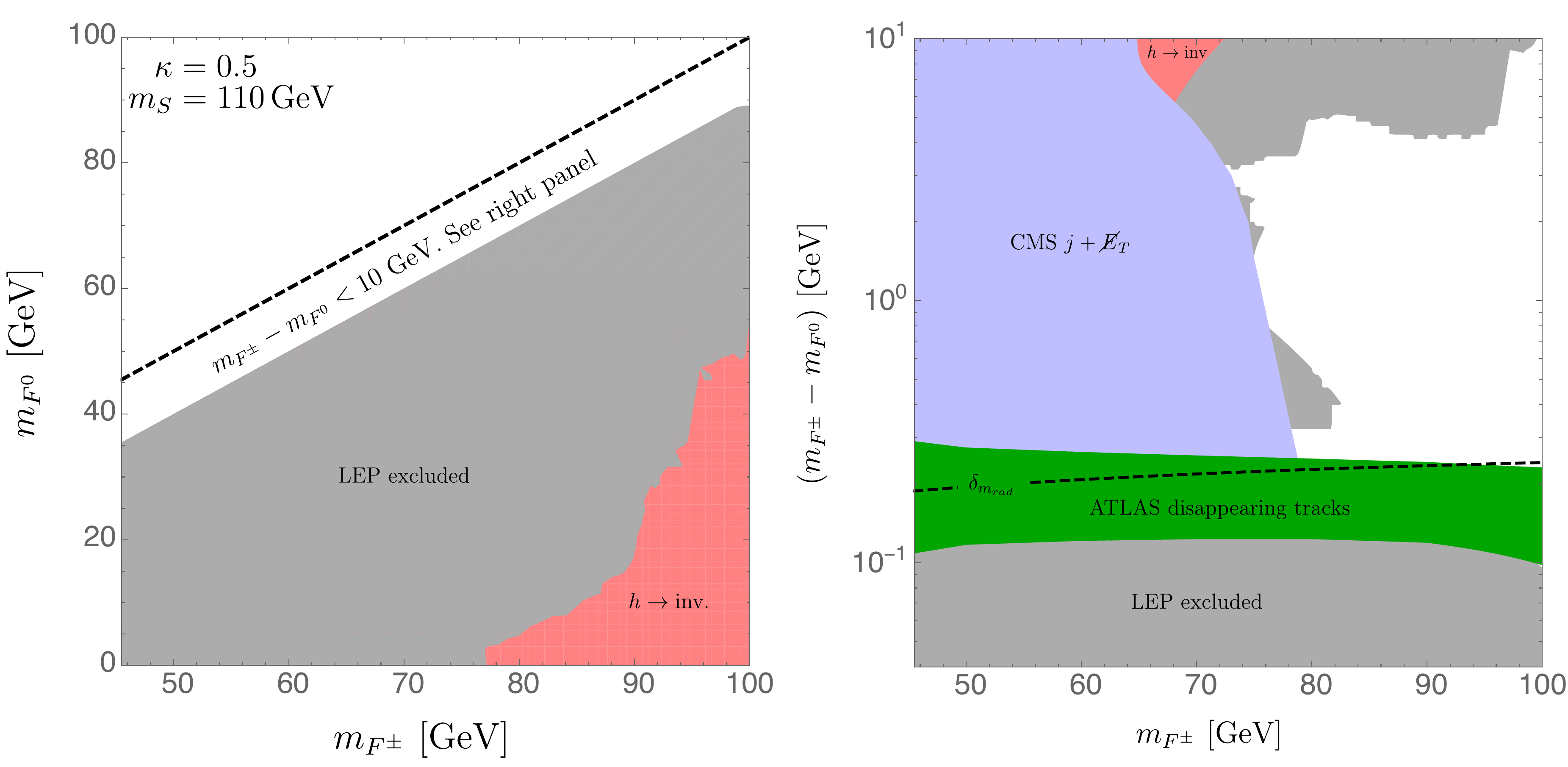}
\caption{Limits on charginos $F^\pm$ (from the simplified model of Sec.~\ref{sec:model}), set by LEP and by the LHC, in the uncompressed region (left) and in the compressed region (right).  The coupling is set to $\kappa=0.5$ and the scalar singlet mass to $m_S=110~\textrm{GeV}$\@.  The LEP limits are the same as in Fig.~\ref{fig:kappa05} but here are indicated in gray, while LHC limits are colored.  In the uncompressed region, the strongest limits are obtained from LEP and from LHC constraints on invisible Higgs decays.  In the compressed region, the strongest limits are obtained from LEP and from LHC monojet and disappearing track searches. The black dashed line indicates the one-loop radiative charged-neutral mass splitting.}
\label{fig:LEPLHC}
\end{center}
\end{figure}

The combined LEP and LHC limits are shown in Fig.~\ref{fig:LEPLHC}, where we indicate with colors the LHC limits, while the LEP limits discussed in Sec.~\ref{sec:leplimits} are shown in gray. 

From the left panel in Fig.~\ref{fig:LEPLHC} we see that the uncompressed region,  $\Delta m \geq 10~\textrm{GeV}$, is completely excluded up to $m_{F^\pm}=100~\textrm{GeV}$ by a combination of LEP results and the constraint on the Higgs invisible width, where in the plot we highlight in red the region which is exclusively ruled out by the Higgs invisible width constraint. 

In the right panel of Fig.~\ref{fig:LEPLHC} we present the limits in the compressed region, $\Delta m \leq 10 \, \textrm{GeV}$\@. In red, blue, and green we show LHC constraints from the Higgs invisible width, monojet, and disappearing track searches, respectively. Some parts of parameter space are excluded by both LHC and LEP, and here we simply overlap LHC constraints on top of LEP constraints. In the uncompressed region, the Higgs invisible width constraints do not lead to any significant improvement with respect to LEP limits, since the couplings of the Higgs to the neutral fermion are suppressed by the small charged-neutral mass splitting (see Eq.~\eqref{eq:ghiggs}). On the other hand, the combination of LHC monojet and disappearing track searches cover the region $m_{\pi^\pm} \leq \Delta m \leq 300~\textrm{MeV}$, which is hard to exclude reliably with published data from LEP displaced searches as discussed in Sec.~\ref{sec:leplimits}. In addition, for $\Delta m \geq 300~\textrm{MeV}$, monojet searches at LHC cover most of the small gaps for charged fermion masses $m_{F^\pm} \lesssim 63~\textrm{GeV}$ left out by our LEP exclusion in Fig.~\ref{fig:kappa05}. 

From Fig.~\ref{fig:LEPLHC} we conclude that the absolute limit on the chargino mass within our simplified model is $m_{F^\pm} \geq 73 \textrm{ GeV}$, and is not improved with respect to the absolute LEP limit. The lower end of this limit is obtained in the compressed region, with mass splittings of a couple of GeV\@. In the uncompressed region, $\Delta m \geq 10 \, \textrm{GeV}$, the combination of available LEP and LHC limits rule out charginos in our simplified model up to at least $m_{F^\pm}=100 \, \textrm{GeV}$\@. The limits are summarized in table \ref{tab:finallimits}.

\setlength\tabcolsep{5pt} 
\begin{table} [h!]
\begin{center}\begin{tabular}{|c||c|c|c|}\hline
 & LEP & LEP+LHC \\ \hline
   $\Delta m \geq 10~\textrm{GeV}$
&  $m_{F^\pm} \geq 76~\textrm{GeV}$
&  $m_{F^\pm} \geq 100~\textrm{GeV}$ \\ \hline
   $\Delta m \leq 10~\textrm{GeV}$
&  $m_{F^\pm} \geq 73~\textrm{GeV}$
&  $m_{F^\pm} \geq 73~\textrm{GeV}$ \\ \hline
\end{tabular}\end{center}
\caption{Combined absolute limits on charginos in the region $m_Z/2~\leq~m_{F^\pm}~\leq~100~\textrm{GeV}$, for the simplified model of Sec.~\ref{sec:model}.  LEP limits are obtained from the searches in Table~\ref{tab:LEPsearches}, while LHC limits consider constraints on the Higgs invisible width~\cite{Aad:2015pla,Khachatryan:2016whc}, disappearing track~\cite{Aaboud:2017mpt}, and monojet~\cite{CMS:2017tbk} searches.}
\label{tab:finallimits}
\end{table}

We conclude the discussion by briefly commenting on the future projected sensitivity from the LHC (with $300~\textrm{fb}^{-1}$).  As shown in Fig.~\ref{fig:LEPLHC}, the remaining window is in the compressed region, where disappearing tracks searches and monojet searches are the most constraining.  The existing disappearing track search already excludes masses up to 100 GeV, but in a limited range of $\Delta m$\@.  The limits from these searches lose sensitivity steeply as a function of decay length, as discussed in section \ref{sec:dtracks}.  Near $\Delta m \approx 300~\textrm{GeV}$, the decay length scales like $\sim (\Delta m)^3$, so that an eight-fold improvement on the lifetime only improves the $\Delta m$ reach by a factor of about 2.  Consequently, extrapolating current searches to $300~\textrm{fb}^{-1}$, we estimate that the limit will improve moderately, by $\sim10~\textrm{MeV}$\@.  Other projections have also been made~\cite{Mahbubani:2017gjh,Fukuda:2017jmk}.

Monojet searches, on the other hand, cover a wide range of $\Delta m$ values, but only extend to $\sim65 - 78~\textrm{GeV}$ in charged fermion mass (depending on $\Delta m$).  A number of projections have been performed~\cite{Schwaller:2013baa, Low:2014cba, Han:2015lma, Barducci:2015ffa, Arbey:2015hca,Baer:2016usl,Curtin:2017bxr,Halverson:2014nwa,Fukuda:2017jmk,Giudice:2010wb} and typically estimate the reach for higgsinos to extend to $\sim 100-200~\textrm{GeV}$\@.  These estimates, however, are strongly dependent on the assumed systematics, making it hard to say conclusively whether or not monojet searches, with the high luminosity data, will be sufficient to cover the remaining parameter space below 100 GeV.

\section{Conclusions} \label{sec:conc}

In this work, we surveyed the limits on charginos with masses ranging from $m_Z/2$ to $100~\textrm{GeV}$\@. We reviewed LEP limits on chargino pair production, and found that charged higgsinos with masses below $\sim 90 \, \textrm{GeV}$ are excluded.  To study limits on fermions with unit charge in a more general scenario, we introduced small modifications to the pure higgsino case in the context of a simplified model.  If a singlet scalar couples to the charged fermions and electrons, then the production cross section is reduced, due to destructive interference, and decay branching fractions are modified.  We showed that for our simplified model, LEP only excludes fermions with unit charge belonging to an $\textrm{SU(2)}_W$ doublet up to $73\, \textrm{GeV}$\@. 

We also discussed LHC limits on such low mass ``charginos''. We discussed a combination of searches, including Higgs precision measurements, monojet, multilepton, displaced decay, and HSCP searches. 
For our simplified model, we found that the LHC, with current statistics, is unable to improve on the overall LEP limit on the mass of charginos.
 The most challenging topology to probe at the LHC corresponds to the compressed region, where charginos decay leaving little-to-no energy deposition in the detector and limits rely mostly on monojet searches.

Our results lead to several questions which remain to be addressed. First, it would be interesting to identify the broader class of models with light fermions with unit charge which are consistent with current data. In this work, we explored charged fermions as part of an $\textrm{SU(2)}_W$ doublet, but a similar analysis may be carried out for other representations. In the case of $\textrm{SU(2)}_W$ singlets,
 fermions with unit charge may evade LEP bounds due to $t$-channel interference in the production cross section, as in this work. 
In the case of $\textrm{SU(2)}_W$ triplets, LEP bounds may also be relaxed with t-channel interference, but we point out that the increased pair production cross section at LHC with respect to the $\textrm{SU(2)}_W$ doublet case should lead to stronger limits from monojet and multilepton searches.  It would be interesting to study the embedding of these models into full UV completions. The case of $\textrm{SU(2)}_W$ triplets is particularly interesting, since it corresponds to the case of the charged wino in the MSSM, where interference in LEP pair production arises through an electron sneutrino. Finally, a careful analysis of the systematics and limit projections at both the LHC and future colliders targeting the low mass region is needed. Future $e^+e^-$ colliders, such as FCC-ee, could definitively test the existence of fermions with unit charge below $100~\textrm{GeV}$\@.

As more data are collected, LHC searches will tend to be optimized for higher mass signals that come into reach.  It is important to be mindful of gaps in exclusion limits, and to identify light particles that are still allowed.  Light particles can serve as a target for future searches, but often require a careful analysis in order to separate from backgrounds.  We have found that charged fermions as light as 75 GeV may have evaded both LEP and the LHC, so far, and therefore serve as a target for future LHC searches.

\section*{Acknowledgements}

The authors would like to thank Kyle Cranmer, Jared Evans, Andy Haas, Philip Harris, Patrick Meade, Carlos Wagner, and Jos\'e Zurita for useful discussions.
The work of D.E.U. is supported by PHY-1620628, M.L. acknowledges support from the Institute for Advanced Study, and J.R. is supported by NSF CAREER grant PHY-1554858.  M.L. would like to acknowledge the Mainz Institute for Theoretical Physics (MITP) and the Aspen Center for Physics, which is supported by National Science Foundation grant PHY-1607611, for their hospitality and support while part of this work was being completed.

\FloatBarrier
\appendix
\section{Validation of LHC Analyses}
\label{app:validation}

\subsection{Monojet Searches}

We validate our monojet analysis by generating events, applying a detector simulation, implementing the monojet selection, and comparing the resulting number of events to the number of events reported by CMS~\cite{CMS:2017tbk}.  The CMS search was performed at 13 TeV and used an integrated luminosity of $35.9~\mathrm{fb}^{-1}$\@.  We compare a sample of $Z(\nu\nu)+j$ events and a sample of $W^\pm(\ell^\pm\nu)+j$ events which have similar kinematics to our signal.  The events are generated at leading order using \textsc{Madgraph5\_aMC@NLO}~\cite{Alwall:2014hca} with MLM matching up to 3 jets, showered with \textsc{Pythia8}~\cite{Sjostrand:2007gs}, and processed through \textsc{Delphes 3}~\cite{deFavereau:2013fsa}.

The ratio between the number of events predicted by our simulation and the number of events found by CMS is defined to be
\begin{equation}
\epsilon_{\rm MC} = \frac{N_{\rm MC}}{N_{\rm CMS}} \, .
\end{equation}
The estimation given by CMS, $N_{\rm CMS}$, is data-driven and therefore accounts for contributions beyond leading order.  In Table~\ref{tab:monojetvalidation} we report the values of $\epsilon_{\rm MC}$ found using the five lowest $\slashed{E}_T$ bins.  Across these bins we find a variation of $8\%$ in $Z(\nu\nu)+j$ and $18\%$ in $W^\pm(\ell^\pm\nu)+j$\@.  The deviation of $\epsilon_{\rm MC}$ from unity by several tens of percent is expected since we generate our events at leading order while the CMS estimation is data-driven so automatically includes contributions from all orders.

\begin{table} [h!]
\begin{center}\begin{tabular}{c|c c|c c}
                             & \multicolumn{2}{|c|}{$Z(\nu\nu)+j$}         & \multicolumn{2}{|c}{$W^\pm(\ell^\pm\nu)+j$} \\ \hline
$\slashed{E}_T~(\textrm{GeV})$ & $~~N_{\rm CMS}~~$ & $~~\epsilon_{\rm MC}~~$ & $~~N_{\rm CMS}~~$ & $~~\epsilon_{\rm MC}~~$ \\ \hline \hline
$250-280$                    & 79700             & $0.68\pm 0.016$          & 49200             &  $0.70\pm 0.03$         \\
$280-310$                    & 45800             & $0.64\pm 0.02$          & 24950             &  $0.76\pm 0.43$         \\
$310-340$                    & 27480             & $0.73\pm 0.027$          & 13380             &  $0.72\pm 0.57$         \\
$340-370$                    & 17020             & $0.64\pm 0.033$          & 7610              &  $0.83\pm 0.082$         \\
$370-400$                    & 10560             & $0.72\pm 0.044$          & 4361              &  $0.88\pm 0.11$          
\end{tabular}\end{center}
\caption{Number of events reported by CMS, $N_{\rm CMS}$, and ratio with the number of events from our simulation, $\epsilon_{\rm MC}$\@.  The reported uncertainties are statistical uncertainties from our simulation.}
\label{tab:monojetvalidation}
\end{table}

\subsection{Disappearing Track Searches}

The ATLAS disappearing track search that we recast was performed at 13 TeV with an integrated luminosity of $36.1~\mathrm{fb}^{-1}$~\cite{Aaboud:2017mpt}.  We parametrize the efficiency to select an event containing a disappearing track by $\epsilon_{\rm event}$ which we factorize, roughly following the parametrization of ATLAS, as
\begin{equation} \label{eq:trkeff}
\epsilon_{\rm event} = \epsilon_{\rm track} \times \epsilon_{\rm selection},
\end{equation}
where $\epsilon_{\rm track}$, the track efficiency, indicates the efficiency to reconstruct a chargino as a disappearing track and $\epsilon_{\rm selection}$, the selection efficiency, indicates the efficiency for the event to be selected.

We compute the track efficiency in Monte Carlo.  The events are generated with \textsc{Madgraph5\_aMC@NLO} for a range of charged fermion masses.  The distribution of decay lengths of the charged fermion is specified by the kinematics of the event and by the charged fermion's lifetime.  In each event, the charged fermions are decayed and assigned a track efficiency taken from Fig.~4 of~\cite{Aaboud:2017mpt}.  Our calculation of $\epsilon_{\rm track}$ is thus a function of charged fermion mass and lifetime.

The event efficiency can be found by comparing the number of signal events produced in our Monte Carlo with the number of signal events needed to produce the limits in Fig.~8 of~\cite{Aaboud:2017mpt}.  With $\epsilon_{\rm event}$ and $\epsilon_{\rm track}$ we use Eq.~\eqref{eq:trkeff} to find $\epsilon_{\rm selection}$, which will also be a function of charged fermion mass and lifetime.

\begin{figure}[htbp]
\begin{center}
\includegraphics[width=8cm]{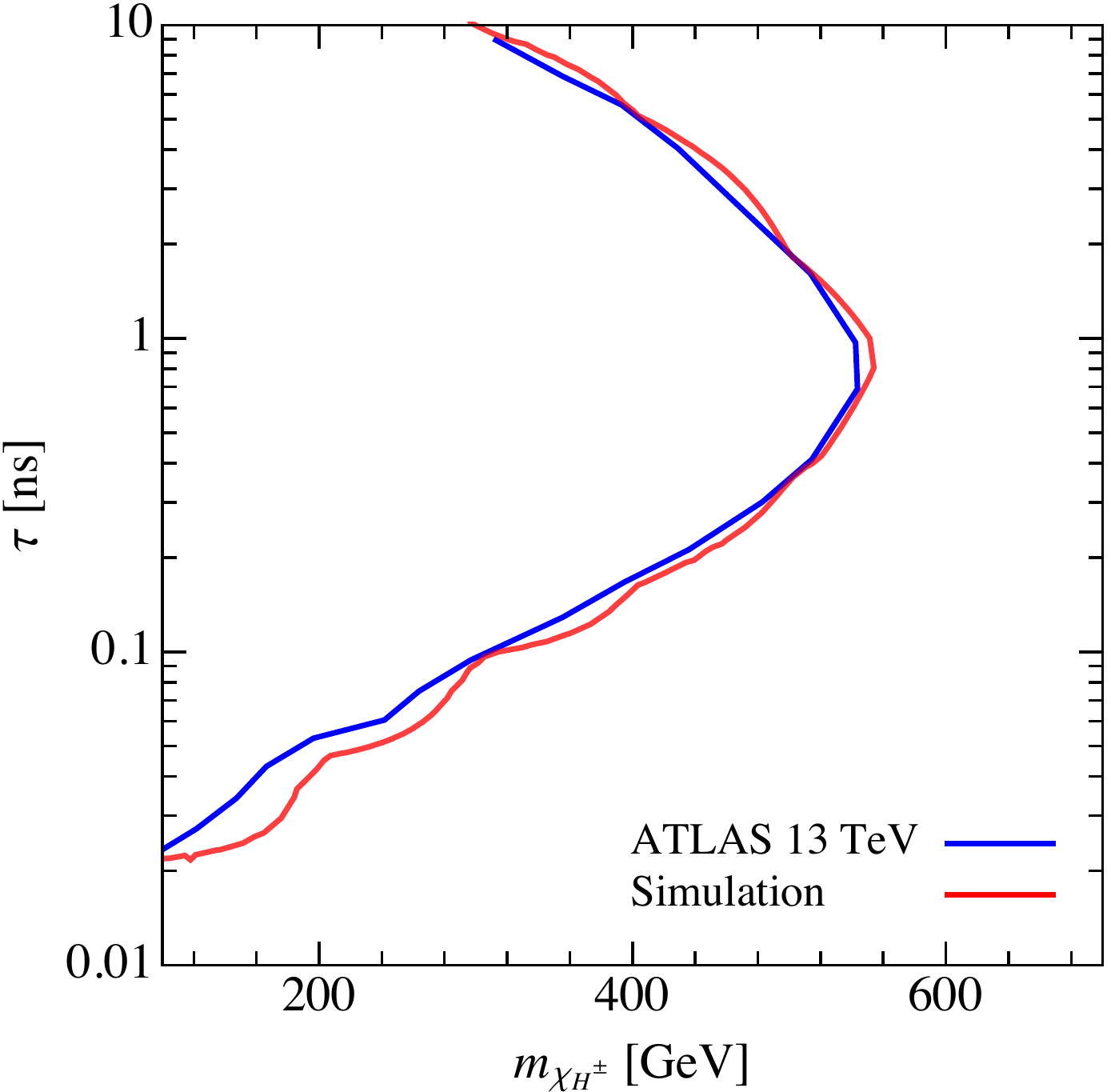}
\caption{Limits from disappearing track search as a function of mass and lifetime from the ATLAS 13 TeV result~\cite{Aaboud:2017mpt} (blue) and from our simulated events with efficiencies applied according to Eq.~\eqref{eq:trkeff} (red).}.
\label{fig:dtracks-validation}
\end{center}
\end{figure}

To compute $\epsilon_{\rm event}$ in our simplified model, we assume that $\epsilon_{\rm selection}$ is the same as in the ATLAS search and compute $\epsilon_{\rm track}$ in simulated simplified model events.  The result of our procedure, applied to chargino events, and the ATLAS result are compared in Fig.~\ref{fig:dtracks-validation}.

\FloatBarrier
\bibliography{treeA_bib}
\end{document}